\relax
\documentclass[a4paper]{article} 
\usepackage{onecolpceurws}  
\usepackage{graphicx}  
\usepackage{tikz}
\usepackage{tabularx}
\usepackage{url}
\usepackage{hyperref}
\usetikzlibrary{arrows,calc,shapes,decorations.pathmorphing,positioning}

\tikzset{
>=stealth',
help lines/.style={dashed, thick},
axis/.style={<->},
important line/.style={thick},
connection/.style={thick, dotted},
vertex/.style={ellipse,draw=black}
}

\title{dAIrector: Automatic Story Beat Generation \\ through Knowledge Synthesis}
\author{
Markus Eger\textsuperscript{1}\ {\normalfont and} Kory W. Mathewson\textsuperscript{2,3}\\
\textsuperscript{1}NC State University, Raleigh, NC, USA\\
\textsuperscript{2}University of Alberta, Edmonton, Alberta, Canada\\ 
\textsuperscript{3}HumanMachine, London, United Kingdom\\
meger@ncsu.edu, korymath@gmail.com
}

\institution{}

\begin{document}

\maketitle

\begin{abstract}
\textit{dAIrector} is an automated director which collaborates with humans storytellers for live improvisational performances and writing assistance. \textit{dAIrector} can be used to create short narrative arcs through contextual plot generation. In this work, we present the system architecture, a quantitative evaluation of design choices, and a case-study usage of the system which provides qualitative feedback from a professional improvisational performer. We present relevant metrics for the understudied domain of human-machine creative generation, specifically long-form narrative creation. We include, alongside publication, open-source code so that others may test, evaluate, and run the \textit{dAIrector}.
\end{abstract}

\section{Introduction}
Improvisational theatre (improv) is an art form in which narratives are developed ad-hoc in front of a live audience \cite{johnstone1979}. Performers are prompted with a concise, ambiguous suggestion (e.g. a location or character relationship) and then share narrative development through action and dialogue. Often these prompts are provided by the audience throughout a performance. The most interesting challenge of improvisation is incorporating new suggestions, seemingly unrelated to the narrative. Improvisation's live justification has been proposed as a model for real-time dynamic problem solving \cite{magerko2009,stein2011improvisation}. Improv has been proposed as a grand challenge for machine learning systems \cite{martin2016improvisational} potentially as an extension to the Turing Test \cite{turing1950computing,mathewson2017turing}. The \textit{dAIrector} collaborates with human improvisors for semi-automated story beat generation, suitable for improvisation performance, through knowledge graph synthesis. First, we describe some background on story generation, improvisational theatre, and plot graphs (from Plotto and TV Tropes). Then, we describe our approach and present quantitative and qualitative evaluation. We conclude with discussion of limitations and future work.

\section{Background and Related Work}

\subsection{Automated Story Generation}
The research problem of automated story generation (ASG) is concerned with generating a sequence which collectively form a narrative \cite{meehan1976metanovel,cook1928plotto}. The sequence can be composed of abstract concepts such as events or actions, or concrete text-based elements such as paragraphs, sentences, words, or characters. Different levels of abstraction and concreteness are accompanied by different challenges. For instance, stories defined at high levels of abstraction maintain step-to-step coherence easier but are simplified and lack unique, specific details.

Previous ASG systems have used symbolic planning and extensive hand-engineering \cite{riedl2010narrative}. Open story generation systems use machine learning techniques to learn representations of the domain from the training data and incorporate knowledge from an external corpus \cite{li2013story}. 
Martin et al. \cite{martin2017event} address the abstraction level challenges by using recurrent neural networks (RNNs) and an event representation to provide a level of abstraction between words and sentences capable of modelling narrative over hundreds of steps. They provide a method of pre-processing textual data into event sequences and then evaluate their event-to-event and event-to-sentence models. Our methods are distinct from 
this technique as we do not focus on the problem of sentence generation from words or characters. The \textit{dAIrector} embraces human co-creators to provide dialog for given plot point descriptions and context. 

Narrative generation approaches, such as TALESPIN \cite{meehan1977tale}, focus on actions, their effects, and element relationships to delineate character intentions \cite{riedl2010narrative} and conflict \cite{ware2011cpocl} which ultimately leads to satisfying an author defined goal. Alternatively, there may be no predefined goal, and systems may discover actions autonomously \cite{theune2003virtual}, ideally resulting in an interesting story. To produce an interesting story in the context of improv theater, however, the prescription of actions (e.g. lines of dialog, character choices, stage directions) is less desirable. A description of a situation suffices to inspire actors who can then translate the prompt into actions, with vague and ambiguous prompts giving the actors freedom to explore the scene \cite{sawyer2003improvised}.

\subsection{Digital Storytelling}

The ad-hoc storytelling experience present in improv theater has been used for research into digital storytelling for more than two decades. Perlin and Goldberg \cite{perlin1996improv} use concepts from improvisational theater to populate virtual worlds, while Hayes-Roth and van Gent \cite{hayes1996improvisational} describe virtual agents that perform improvisational theater, modifying their appearance to convey simulated emotional state. Several knowledge-based approaches have been proposed for various problems in the space of acting in the improv theater, such as scene introduction \cite{o2011knowledge}, fuzzy reasoning \cite{magerko2011employing}, affect detection \cite{zhang2007affect}, and robotic actors \cite{mathewson2017improvised}.

Through collaboration between human and machine, complex stories can be constructed. Generative plot systems have been developed for nearly 100 years \cite{cook1928plotto}. These systems aim to aid human creators with algorithmically generated prompts to explore diverse plots. Through interaction with generative systems, users are inspired to engage with topics they would not have otherwise. The excitement of exploring unknown spaces and engaging with novel topics defines the artform of theatrical improvisation \cite{johnstone1979}.

\subsection{Improvised Theatre}
In improvised theatre (improv) there is no script and no rehearsal; the show is written and performed at the same time. It is an art derived from the spontaneous justification of pseudo-randomness. Improvised theatre has been described as a suitable test bed for human-machine co-creation systems \cite{magerko2009,mathewson2017improvised,martin2016improvisational}. In improv, performers must attend to, and remember, details in the story and must synthesize previous information with novel dialogue and actions to progress a narrative. Often, the use of external prompts (or suggestions) are utilized to add entropy to the performance \cite{johnstone1979}. This motivates the actors to justify this information within the context of the current scene \cite{sawyer2003improvised}. 

Improvised scenes can be summarized in three stages: platform, tilt, finding a new normal \cite{johnstone1979}. The \emph{platform} of the scene defines what is normal in the universe (i.e. who, what, when, where). The \emph{tilt} provides flavour\footnote{According to www.thewayofimprovisation.com/glossary.php: \emph{A tilt re-frames the scene with a different context.}}. It is what makes this particular performance unique from others with a similar platform. Finally, \emph{finding a new normal} is how the scene justifies the tilt towards resolution. These three stages enable investigation of the ability of the \textit{dAIrector} to generate cohesive plots and develop context-rich narrative. Our work addresses the specific aspect of 
generating prompts for the actors on stage during an improvised theater performance. These prompts constitute
the beats of the story in form of a platform, as well as tilts for the actors. It is then up to the actors
to act out the scene to find a new normal.

\section{Approach}

We present an improvised narrative co-creation system called \textit{dAIrector} which acts as an automated director to collaborate with humans storytellers for live improvisational performances and writing assistance. The generated stories are represented as linked clauses taken from William Wallace Cook's ``Plotto: The Master Book of All Plots'' \cite{cook1928plotto} augmented with related information from TV Tropes\footnote{\url{http://tvtropes.org/}}---a wiki-style database that contains narrative tropes occurring in a wide range of different narratives. Human artists can rapidly link the provided prompts to collaboratively evolve a narrative through dialogue and actions. In this way, the \textit{dAIrector} augments human creativity. We discuss the challenges of evaluating a tool that, by design, provides ambiguous guidance. We conclude by presenting several directions for future research.

\subsection{Plotto}
Our work builds on the narrative development book ``Plotto: The Master Book of All Plots'' by William Wallace Cook \cite{cook1928plotto}, which contains a large variety of plots. What makes it suitable for a computational application is the graph structure. Rather than enumerating plots, Cook split them into fragments with instructions on how to combine them. 

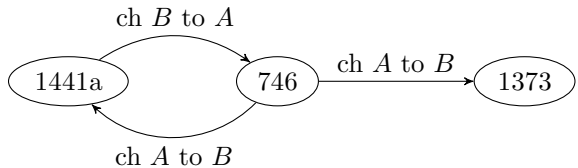
\begin{figure}
\begin{center}
\begin{tikzpicture}[
        scale=1,transform shape,
        every node/.style={color=black}
    ]
    \node (n746) [vertex] {746};
    \node (n1441a) [vertex, left of=n746, xshift=-1.75cm] {1441a};
    \node (n1373) [vertex, right of=n746, xshift=2.25cm] {1373};
    \draw [->] (n746) to [bend left=45] node[midway, below] {ch $A$ to $B$} (n1441a);
    \draw [->] (n746) to node[midway, above] {ch $A$ to $B$} (n1373);
    \draw [->] (n1441a) to[bend left=35]  node[midway, above] {ch $B$ to $A$} (n746);
\end{tikzpicture} 
\caption{A sub-graph from Plotto. Nodes are plot fragments with corresponding numbers from Plotto, edges are connections between plot points and edge labels correspond to instructions for character modifications in subsequent fragments.}
\label{fig:plotto}
\end{center}
\end{figure}

The plot fragments constitute nodes and edges between them describe which fragments can be connected to obtain a story. Edges can have labels, which contain instructions for changing character symbols in subsequent plot fragments (e.g. changing character $A$ to character $B$). Figure  \ref{fig:plotto} shows a subgraph from Plotto, each node a plot point and each edge a modification. Fragment $746$ is defined in Plotto as: \emph{B, who was thought by the people of her community to have supernatural powers, is discovered to have been insane - a condition caused by a great sorrow}. This fragment can be followed by either fragment $1441a$: \emph{A seeks to discover the secret of Life}, or fragment $1373$: \emph{A sells his shadow for an inexhaustible purse}. Both of these fragments make no mention of $B$, who is the main character of fragment $746$. Therefore, the modifying edge instructs us to change $A$ to $B$, ensuring consistency of the characters used. These three nodes represent just a small fraction of the entire $3000$ notes contained in Plotto. We automatically parse the nodes and edges into a JSON-based representation of the graph. Thus, generating a story is done by performing a walk through the graph starting at any random node.

Eger et al. used a similar method to build a plot generator utilizing the plot fragments \cite{eger2015plotter}. Since the plot fragments in Plotto are abstract descriptions of plot points, ambiguous, and contain symbolic names for the characters, the generated plots are less than suitable for presentation to an audience. The ambiguity and openness make them ideal for interpretation in improvised theatre. These plot points represent platforms for scenes additional related details are needed to tilt the scene.


\subsection{TV Tropes}
TV Tropes is a wiki-style website that contains narrative tropes, i.e. patterns or situations that occur across a variety of different narratives. As a wiki, tropes often contain references to other, related tropes. Of particular interest for our work are TV Tropes's \textit{plot tropes} which describe high level plots abstractions. While related to those plots in Plotto, this TV Tropes graph contains unlinked semantically related, complimentary information. 

The story beat is an identifiable moment of change in a narrative \cite{mckee1997substance, mateas2003faccade},
either in the 
form of a new platform or as a tilt. By connecting the Plotto graph plot points as platforms and the TV 
Tropes plot tropes as tilts, the \textit{dAIrector} creates complete abstract narrative descriptions. 

As described above, story generation using Plotto can be thought of as a walk through the graph of plot fragments. For a performance, \textit{dAIrector} starts at a random node in the graph by presenting it to the actors on stage. We call the plot fragment presented to the actors by the system the \emph{platform}. The actors can prompt the system for 1) the next platform in form of a new plot fragment or 2) for a tilt to refine the current scene, and the system will use the platform to determine which plot fragment or tilt to present next.

\subsection{Paragraph Vectors}

Our system utilizes paragraph vectors to provide information dependent on the current platform.
Paragraph Vector is an unsupervised machine learning method to represent variable-length input text in 
a dense, fixed-length feature vector \cite{le2014distributed}. Paragraph vectors overcome 
limitations of ignoring word order and semantics of naive bag-of-word methods. To train the paragraph 
vector model, we use the full text of the plot fragments, as well as the descriptions of the tropes. 
We follow the training as described by Le and Mikolov and use the Doc2Vec method from Gensim 
\footnote{\url{https://pypi.org/project/gensim/}, \url{https://radimrehurek.com/gensim/models/doc2vec.html}}. 
Training parameters are as follows:
\begin{itemize}
    \item Dimensionality of feature vector: $410$
    \item Initial learning rate: $0.03$
    \item Maximum distance between the current and predicted word within a sentence (window): $4$
    \item Ignore all words with fewer than two occurrences
    \item Negative sampling is used with $4$ noise words
\end{itemize} 
All other parameters are defaults as defined in Gensim.

During a performance, the current platform is used as to find the next platform or a tilt.
Alternatively, instead of using the current plot fragment, actors may also provide a custom prompt 
to the system to steer the plot in a certain direction. When queried, the \textit{dAIrector} returns 
$5$ tilt options with minimum cosine distances in vector space from the entire space of candidate 
options \cite{mikolov2013distributed,cer2018universal}. For vectors $\mathbf{A}$ and 
$\mathbf{B}$ the cosine distance, or angle distance between two vectors, is defined as 
$d(\mathbf{A}, \mathbf{B}) = 1 - \frac{\mathbf{A} \cdot \mathbf{B}}{\left\Vert \mathbf{A} \right\Vert \left\Vert \mathbf{B} \right\Vert}$.

As noted above, we use the full text of the plot fragments, as well as the descriptions of the tropes
to train the paragraph vector model. However, since these descriptions are typically several 
paragraphs long, we only communicate the trope names to the actors. While we perform all 
comparisons against the full textual description of the tropes, this description is never displayed 
to the actors. This allows us to keep the instructions from the \textit{dAIrector} concise open 
ended while providing additional, related information \cite{sawyer2003improvised}.

\subsection{Plot Tree Generation}
Rather than generating a single plot from the Plotto graph we generate a tree, with one start node chosen 
randomly and all successors as children. Each of these children has its successors as children up to a 
configurable depth. By default, the platform used to determine the next scene is the current plot fragment, 
but actors may prompt the system for a plot fragment that aligns more closely with their interpretation of 
a scene or details which arose from the scene improvisation. A performance of such a plot tree starts
at the root node, and proceeds down the tree, where child nodes are chosen depending on the prompts given
by the actors. 

For example, the plot fragment \emph{Carl's friends, Doug and Fred, believe that Lisa, whom Carl is about to marry, is a woman of immoral character}, has two successors: \emph{Carl seeks to free himself from certain meddlesome influences} and \emph{Lisa, harassed by gossip that reflects on her integrity, seeks deliverance from false suspicion}. Depending on which aspect of the original plot the actors decide to focus on, the interference of Doug and Fred in Carl's affairs or the rumors that Lisa is of immoral characters, one or the other successor is better suited. Note that neither successor is completely unsuitable in any case, meaning that the platform can provide guidance for the system, but that guidance does not necessarily have to be followed.

\subsection{Contextual Tilts}
In addition to simply traversing the plot tree, the system can also provide \emph{tilts} in the form of plot tropes obtained from TV Tropes. In this case, the platform is used to find a selection of tropes that fit best with the current plot fragment. Tilts provide additional information. Thus, rather than returning the single trope is closest to the platform, our system computes the $5$ most semantically similar tropes and then returns a random sample from the related set.

Prior to sampling, plot fragments are filtered and excluded if they only provide redundant information. For example, if the platform is \emph{Albert, an inefficient, futile sort of person, comes to believe that he is the reincarnation of Nicola Tesla}, the best fitting trope according to our system is \emph{Reincarnation}, which does not provide any additional information. However, the trope that has the second lowest distance is \emph{Deal With The Devil}, which provides additional guidance for the actors. In a performance, the actors can utilize this, for example, to narrow down how Albert came to his belief, by making a deal with the devil. This provide the scene additional directions to explore, but it is up to the actors to decide when they would prefer a tilt for additional guidance and when they want to follow their own impulse for where the scene should go. 

To eliminate redundant tropes, we compute the word intersection of words with more than $3$ letters (to exclude articles, pronouns, etc.) with the platform and discard any tropes for which this intersection is non-zero. In the example above, because the word \emph{reincarnation} is also part of the platform this particular tilt would not be provided by the system.

\subsection{Stage Presence}

For a performance, the system provides output in the form of platform beats and tilts, according to prompts
given by the actors. The basic outline of this process is as follows:

\begin{verbatim}
platform = root(plot_tree)
present(platform)
while platform != Null:
    request, prompt = get_input()
    context = platform
    if prompt != Null:
        context = prompt
    if request == "platform":
        platform = best_match(context, children(platform))
        present(platform)
    if request == "tilt":
        tilt = random(best_n_match(context, tvtropes, 5))
        present(tilt)
\end{verbatim}

When the actors request either a platform beat or a tilt, the system uses the paragraph vector model to 
find the best match among all candidates for the given prompt, which defaults to the last presented 
platform beat. For the next platform the candidates are the Plotto plot fragments that are children
in the provided plot tree, while for tilts the candidates are all TVTropes plot tropes. 

The basic mode of interaction with our system is through a console-based application. This application will present the plot fragments in order, and can be prompted for the next platform or a tilt. For a live performance, this mode is less convenient and therefore we also provide the capabilities for speech input and output, realized through speech-to-text using \emph{pocketsphinx} \cite{huggins2006pocketsphinx} and text-to-speech using the built-in \texttt{say} operating system command. Using this interaction mode, the system reads plot fragments and tilts out loud, and the next platform beat or tilt can be obtained by
the actors saying the corresponding keywords.

Plot fragments, as contained in Plotto refer to characters in the story using codes, including $A$ for the main male character, $B$ for the main female character, but also very specific codes such as $AUX$ for a fictitious aunt. As part of presenting the plot fragments to the actors and audience, our system replaces these codes with consistent character names. This is controlled via a configuration file, with default names provided by the system. For clarity, we present all plot fragments in this paper with names replaced. Note that at this point
we do not change pronouns if character symbols are replaced since that would require automatic identification
of the referent of each pronoun, which is outside the scope of this work.

\section{Evaluation}
Given the unique environment of an improv theater performance, evaluating the quality of output is challenging; there is often no catastrophically ``wrong'' output \cite{mathewson2017improvised}. That said, given the platform some tilts will require significantly more justification to produce a satisfying narrative \cite{sawyer2003improvised}. For our work, the main challenge lies with evaluating the platforms and tilts of story fragments. 


Having humans annotate multiple story fragments with the best fitting trope is a challenging multi-class classification problem. For example, the plot fragment \emph{Joe, becoming aware of an old prophecy, unconsciously seeks to become like the exalted protagonist of the prophecy} could be seen as having any of the tropes \emph{Prophecy Twist}, \emph{Self Fulfilling Prophecy}, \emph{The Unchosen One}, or \emph{Because Destiny Says So} (among others) as the ``correct'' fit. For this task, we report how well our system reproduces human-assigned tilts on our test set.

\subsection{Evaluation of Tilts}

To test the key functionality of \textit{dAIrector}, that of selecting a best-fitting tilt given a 
plot fragment, we design a simple task. First, we generated a dataset of $100$ clean, labelled tilt -- 
plot fragment pairs. We split this dataset into training and testing sets. For evaluation, we sample 
a random plot fragment from the test set and the task for the system is to correctly predict the associated tilt. Given a plot fragment our model returns the 5 candidate tilts with the minimum 
cosine distance. We evaluate the system based on top-5 accuracy. Clearly stated for explicitness, 
given a plot fragment, how likely is it that the associated tilt is in the $5$ results returned by 
the system? This number is reported over the entire $20$ examples in the test set.

One insight gained from this approach was that while the tropes produced by the system are usually related to the text fragment, there are several tropes that are vague and apply to many scenarios, while others, which are often more closely related to the story fragment at hand are more specific. For example, in our test set, the trope ``Much Ado about Nothing'', which is a generic trope about love, applies to a wide variety of plot fragments. The trope that was selected the second most often for our test set was, ``Road Trip'', which applies to a wide variety of travel-related scenarios.


The \textit{dAIrectory} returns a random sample from the five most closely related tilts to a given plot point. It is therefore also reasonable to use the top-5 error as a measure of quality rather than the top-1 error. Top-N error is a common error metric for classification tasks and measures how often the target class does not show up in the top-N classes assigned to a test example. 

Even so, the top-5 error on our test set is $40\%$, while the top-1 error rate is $66\%$. While high, the 
trope annotation task resulted in many arbitrary choices by human annotators. Most likely this is due to 
there not being a clear best trope, and human annotators being overwhelmed by the number of possible, subtly 
different, tropes. 

Our plot trope set contained about $700$ tropes from the total trope set of about $4300$\footnote{These are 
the $700$ tropes listed as \emph{plot tropes} by TV Tropes:
\url{https://tvtropes.org/pmwiki/pmwiki.php/Main/Plots}
}. 
For example, the story fragment \emph{Alfred is thrown into prison through false evidence in a political conspiracy} was assigned the trope \emph{Get Into Jail Free} by a human annotator, but our system returned \emph{Clear Their Name}, \emph{Mystery Literature}, \emph{No Mere Windmill}, \emph{Lipstick Mark}\footnote{as in evidence of a cheating spouse.}, \emph{Prison Riot} as the top five tropes, all of which could also be deemed applicable. 

Note the difference in specificity between \emph{Mystery Literature} and \emph{Lipstick Mark}, where the latter provides a lot more detail to the actors of how to proceed with the scene. At present, our system treats all tropes as equally applicable, but, as noted above, some tropes are more general and thus related to the story being presented while actually adding less detail than others. While very specific, the definition of the trope \emph{Get Into Jail Free} actually refers to a character that wants to get arrested intentionally, which is arguably less fitting with the given sentence, but demonstrates the challenges faced by the human annotators.

As a way to quantify this discrepancy we used information contained within the TV Tropes graph. Tropes linked from one another if they shared some commonality. We used these links to calculate a distance between tilts our system generated and the humans annotated, equal to the number of links between two tropes. 

For example, \emph{Get Into Jail Free} links to \emph{Can't Get in Trouble for Nuthin'}, which links to \emph{FrameUp} as a reason for the arrest. That trope links to \emph{Clear Their Name} as a way to resolve the situation, resulting in a distance of $3$ between \emph{Get Into Jail Free} and \emph{Clear Their Name}. 


Over the entire TV Tropes set, the median distance between two tropes is $3$ (mean: $3.1$, stddev: $0.6$). This is not unexpected, since tropes often refer to ``supertropes'', which refer to other ``supertropes'', from which the target trope can be reached. However, the median distance for tropes given as tilts by our system, excluding those that exactly matched the human annotation was $2$ (mean: $2.5$, stddev: $0.7$), which is typically a connection via the ``supertrope'' common to the two tropes. 

\subsection{Sample Stories}

To better illustrate the output from our system, we present sample stories produced by the system along with tilts the actors might be given. These stories demonstrate how the produced outputs are coherent stories. The coherency is a product of the structure employed by Plotto. Additionally, we show examples of stories to illustrate that the tilts the system selects can be used to refine, provide background information, or drive the story in different directions. To demonstrate our system, we set the maximum length of each story to $5$. Shorter stories tend not to have enough happening in them to qualify them as stories, while longer stories, owing to the structure of the book, start to meander to different, somewhat unrelated story lines. 

The first example is a well structured narrative with a beginning, middle, and end, consisting of the following
platform beats:
\texttt{
    \begin{enumerate}
        \itemsep0em 
        \item Lana, a person influenced by obligation, falls into misfortune through mistaken judgment. 
        \item Lana, in order to be revenged upon her enemy, Mr. Kyle, manufactures an infernal machine, BLOB. 
        \item Lana, influenced by a compelling idea of responsibility, finds it necessary to protect her friend, Tynan, from a secret danger.
        \item Lana, suspected of treachery by her friend, Tynan, in a daring rescue, saves the property and perhaps the life of Tynan, and proves her faithfulness by a revelation of the danger to which Tynan, unknown to himself, was exposed. 
        \item Lana seeks to correct a character weakness in her friend, Tynan.
        \item Lana achieves success and happiness in a hard undertaking.
    \end{enumerate}
}

In each scene, the platform is clear and evident. In addition to the platform of the scene and the dialog from the improvisors, the actors might desire a plot device to instigate or inspire the action. For instance, the "secret danger" referenced in Scene 3 is vague, and the actors might ask the system for a tilt. One such applicable tilt returned by our system, \emph{It Belongs in a Museum}, provides context to further refine the ``secret danger''.

Plot fragments present in Plotto rarely mention time passing, and it is often up to the actors to explain jumps in time. Consider this example of a sequence of platform beats from our system:
\texttt{
    \begin{enumerate}
    \itemsep0em 
        \item Alfred, in order to restore to Beatrice, without a confession of culpability, wealth of which he has secretly defrauded her, marries her.
        \item Alfred seeks to escape difficulties, restore property and be free of an unloved wife, Beatrice, all by secret enterprise.
        \item Alfred leaves his coat on a cliff at the seaside, drops his hat in a stunted tree below the brink, and vanishes from the scenes that know him.
        \item Alfred, under a fictitious name, returns to his native place, where he had committed a youthful transgression, and, as an Unknown, seeks to discover  whether his youthful escapades have been forgotten and forgiven. Also, he wishes to make reparation in an assumed character for wrong done in his true character.
        \item Alfred, returning as an Unknown to his native place, discovers no one recognizes him.
    \end{enumerate}
}

Alfred disappears in scene 3, and then reappears what is apparently much later in scene 4 to wrap up his transgressions from the earlier scenes. We highlight this story, because when the system was asked for a tilt on scene 3, it responded with the tilt \emph{Tailor Made Prison}, which seemed unfitting at first. However, upon reflection the references to the coat and hat link it to the concept of a tailor. This is remarkable for two reasons: 1) it is probably not a connection that would arise immediately to a human, and 2) this can be seen as a pun that works well within the context of our domain.

\subsection{Qualitative Evaluation by Professional Improvisor}

To investigate the quality of the system we instructed a professional improvisor (fluent, native English speaker, improvisor with 10+ years performance experience) to interact with the \textit{dAIrector}. The performer was given an introduction to the system, and then explored the interaction over the course of several scenes. The performer discussed their impressions during the interaction. We summarize the interaction feedback below by including \emph{quotes from the performer}. For several points, we directly address the quotes.

\begin{itemize}
   \itemsep0em 
    \item \emph{There is a real fun in getting yourself into trouble and then putting your faith in the \textit{dAIrector} to do something to help.} In improvisation these are described as What Should I Do? moments, when the improvisor decided to prompt the system for the next plot point or tilt.
    \item \emph{It doesn't know what I want the scene to be about or what decisions I make}. This is an area for future work focused on how the actor's dialogue and actions are incorporated as prompts for the system.
    \item \emph{Justification is natural, and it is natural to make leaping assumptions to connect actions/intentions to characters as the plot generation system did not inherently make those assignments.}
    \item \emph{Tilts don't over-complicate the narrative, it expands the story rather than advancing the plot, it adds flavour. The tilt is not always necessary, and making it optional is suitable in improvisation. That said, I prefer to use the system with the contextual tilts. They run the risk of throwing a curveball that is very difficult, but they are often the sort of thing that would be very fun to play. Tilts are a fun expansion.}
    \item \emph{I prefer being able to choose when the plot points and hints come} 
\end{itemize}

\section{Discussion and Conclusions}

One of the limitations of our approach is that in our graph representation of domains all nodes are created equal, even when the underlying data sets might have additional information attached to them. For example, the tropes in the TV Tropes data set actually frequently refer to ``subtropes'', ``supertropes'' or  even state ``contrast with'' or ``opposite of'' in relation to another trope. Our system often returns very broad tilts such as \emph{Mystery Literature}, or overly specific tilts such as \emph{Lipstick Mark}, without any means to control which one to get. However, we believe that utilizing the structural information contained within the data set could lead to tilts that are better suited for any application. Extracting this structural information is non-trivial as it is not structured meta-data. Additionally, while TV Tropes is a useful resource, it is a large dataset that suffers from common dataset quality and inconsistency issues. One way to address these limitations would be to use a subset of tropes that are particularly narrative building. This would require filtering based on a heuristic (learned or pre-defined) which can classify and rank tropes based on narrative building qualities. Some of these qualities could be information about the universe gained through introduction of the trope, or dynamic shifts between characters

We used Plotto and TV Tropes as our data sources because they cover a wide range of different narratives. It would also be possible to expand the \textit{dAIrector} to use more specialized databases such as DramaBank  \cite{Elson2012DramaBankAA}. By designing a structured graph of information,  textual plots of TV show episodes could constitute an interesting data source as well. This could allow for plots which extend over seasons, character arcs, individual episode, or scenes within an episode.

Treating the plot fragments as a graph allows us to use the story-generating walk for other data sources which can be represented as graphs. We are also considering a possible application outside of narratives: the directed exploration of large graphs, such as Wikipedia for knowledge synthesis. In this scenario we could target users browsing a certain topic, starting an article of interest. The user could then request linked articles, constrained to articles semantically related to a query from a different data source.

\section{Acknowledgements}
We would like to thank the anonymous reviewers for their insightful feedback and suggestions toward improving this research. Additionally, we would like to thank several individuals for their invaluable discussion in the preparation of this work, they include: Piotr Mirowski, Stuart Hoye, and Gary Ka\u{c}mar\u{c}\'ik. Finally, we are indebted to the talented perfomers at Rapid Fire Theatre who provided expert opinions on the system, including: Paul Blinov and Julian Faid. 

\bibliographystyle{alpha}
\bibliography{bibliography}
\end{document}